\documentclass[proof,dvipdfmx]{pasj01}
\Received{$\langle$reception date$\rangle$}
\Accepted{$\langle$acception date$\rangle$}
\Published{$\langle$publication date$\rangle$}

\begin{document}

\title{Spherical explosion with central energy source}
\author{Miyu Masuyama\altaffilmark{1,2}, Toshikazu Shigeyama\altaffilmark{1,2}, and Yoichiro Tsuboki\altaffilmark{1,2}}%
\altaffiltext{1}{Department of Astronomy, Graduate School of Science, University of Tokyo, Bunkyo-ku, Tokyo 113-0033, Japan} 
\altaffiltext{2}{Research Center for the Early Universe, Graduate School of Science, University of Tokyo, Bunkyo-ku, Tokyo 113-0033, Japan}
\email{masuyama@resceu.s.u-tokyo.ac.jp}


\maketitle

\begin{abstract}
We present a novel semi-analytic solution describing the propagation of a spherical blast wave driven by a central energy source. The initial density profile has a power-law function of the distance from the center and  the energy is injected only into the central region at a rate given by a power-law function of time. This solution is composed of three regions separated by the contact surface and the shock front. The innermost region is assumed to be uniform and the outside of the contact surface includes the shocked matter described by self-similar solutions. 
We analytically derive the applicable range of parameters of this solution from requirements to satisfy boundary conditions. We perform numerical simulations for flows with various values of parameters,  some of which reside out of the thus derived applicable range, and compare with the semi-analytic solutions. 
\end{abstract}

\section{Introduction}
In many astronomical events, a transient flow of gas often exhibits self-similarity. In fact, self-similar solutions are known for specific phenomena. The examples include gravitational collapse of the core of a massive star at the end of its evolution \citep{1980ApJ...238..991G,Yahil1983}, blast wave propagating inside the star as a result of the core bounce, emergence of a shock from the surface of a star \citep{Sakurai1960, NakayamaShigeyama2005, SuzukiShigeyama2007}, blast wave featuring a supernova remnant consisting of shocked ejecta and circumstellar matter (CSM) \citep{Chevalier1982}, and blast wave caused by a point explosion inside matter with the power-law density distribution \citep{Sedov1959, Taylor1950}. \citet{Byung1998} generalize Chevalier's self-similar solution to include a contribution from a pulsar activity. \citet{BlandfordMckee1976} presents a self-similar solution for ultra-relativistic flow as a result of point explosion with a central energy source whose energy injection rate follows a power law with respect to time.  \citet{Ostriker1988} thoroughly reviewed self-similar solutions applicable to astrophysical phenomena including the above mentioned solutions. Here we will discuss the corresponding non-relativistic flow using the Sedov-Taylor self-similar solutions.

 \citet{Dokuchaev2002} presents a self-similar solution for spherically symmetric blast wave with an energy source, in which the term of the energy injection rate is introduced in the energy equation as a power-law function of time. Thus the energy is uniformly injected into the entire flow independent of the radius. Though this solution can be used to describe the dynamics of radio-active matter ejected from a neutron star merger, this form of energy source might limit the range of applications. Spherical blast wave is generated by many kinds of central energy sources such as neutrino emission from a proto-neutron star, rapidly rotating magnetar, and other kinds of activities of the central compact objects.  \citet{SuwaTominaga2015} treated a spherically symmetric blast wave driven by the central energy source using the thin shell approximation. Thus their model does not describe the structure inside the wave. Though \citet{Ostriker1988} treated several astronomical situations for a blast wave driven by an  energy source, they did not describe the location of the energy source or the structure of flows. 

In this paper, we describe spherical blast waves driven by a central energy source with an energy injection rate given by a function of time $t$ measured from the onset of the energy injection as
\begin{equation}\label{eq:eneinject}
\dot{E}(t)=At^{-k},
\end{equation}
where  positive constant parameters $A$ and $k$ have been introduced.
 The blast wave propagates in the ambient matter with a density profile
 \begin{equation}\label{eq:rhopro}
 \rho_0(r)=Br^{-\omega}, 
\end{equation}
where $B$ and $\omega$ are positive constants and $r$ is the distance from the center. We can make a unique non-dimensional quantity from these dimensional constants $A$ and $B$ together with the independent variables $t$ and $r$. Using this non-dimensional variable, we construct a self-similar solution for the blast wave equivalent to the Sedov-Taylor solution. In general, these solutions have contact discontinuities at finite radii and do not reach the center. Using this feature, we construct a model in which  a blast wave driven by the central energy source embedded in an ultra-relativistic gas or photon gas pushing the outer region through the contact discontinuity. We check whether the derived solutions are realized by performing numerical simulations. We constrain the range of non-dimensional parameters $k$ and $\omega$ for which self-similar solutions exist. Furthermore, we numerically investigate blast waves generated with these two parameters for which a self-similar solution does not exist.

The following is the outline of this paper. In Section 2, we introduce the setting of our model. In Sections 3 and 4,  the procedure to derive generalized self-similar solutions, numerical solutions of hydrodynamical equations are introduced, respectively. In Section 5, the self similar solutions are compared with numerical solutions and we discuss the behavior of waves especially near the contact surface. Section 6 summarizes and concludes the paper.

\section{Model}\label{sec:model}
As already mentioned in the previous section,  our model describes flows driven by a central energy source characterized by equation (\ref{eq:eneinject}) propagating in the ambient medium with the density profile given by equation (\ref{eq:rhopro}), while \citet{Dokuchaev2002} used a uniform density profile. It should be noted here that the non-dimensional quantity becomes identical to that of the original Sedov-Taylor solution if $k=1$, though the corresponding solution does not describe a prompt explosion. Integrating the energy injection rate in equation (\ref{eq:eneinject}) from $t=0$ to $t=\tau$, we obtain a positive and finite amount of energy
\begin{equation}
E=\int_0^\tau \dot{E} dt=\frac{A}{1-k}\tau^{1-k},
\label{eq:totalinject}
\end{equation}
 if $0\leq k<1$.

We consider spherically symmetric flows governed by the following equations.
\begin{eqnarray}
\label{eq:mo}
\frac{\partial \rho}{\partial t}+\frac{\partial (\rho v)}{\partial r}+\frac{2\rho v}{r}=0, \\
\label{eq:con}
\frac{\partial v}{\partial t}+v\frac{\partial v}{\partial r}+\frac{1}{\rho}\frac{\partial p}{\partial r}=0, \\
\label{eq:ene}
\frac{\partial}{\partial t}\left(\frac{p}{\rho^{\gamma}}\right)+v\frac{\partial}{\partial r}\left(\frac{p}{\rho^{\gamma}}\right)=0,
\end{eqnarray}
where $\rho,~v,~p$ are  the density, the velocity, and the pressure, respectively. They are functions of  $r$ and $t$. The constant $\gamma$ denotes the adiabatic index.

\begin{figure}[t]
\begin{center}
\includegraphics[clip,width=10.0cm]{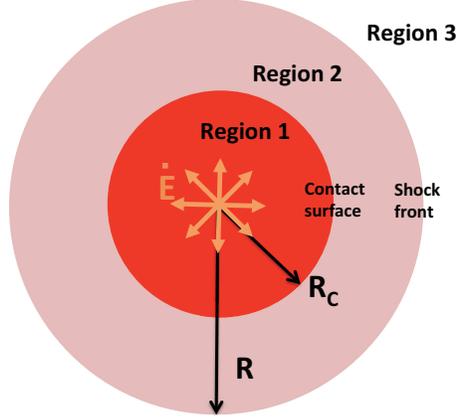}
  \end{center}
  \caption{The schematics of our model.}  \label{fig:1}
\end{figure}

The self-similar solution with a central energy source or inhomogeneous ambient matter does not reach the center and has a contact surface in between. Accordingly the gradient of the density diverges at the contact surface.  Thus we divide the flow into three regions as shown in Figure 1. The self-similar solution describes the flow in region 2. To connect this self-similar flow with the central energy source, we introduce  region 1 inside the contact surface.

We assume that  region 1 is very hot and dominated by radiation and that the density of gas is negligible there because the gas is swept up by the blast wave and concentrated in region 2.
This region is assumed to have uniform profiles of physical quantities and evolve following the first law of thermodynamics. 
The energy conservation indicates that the injected energy is converted into two components;  the internal energy $E_1$ and the work done by the photon gas to the post shock fluid.
Therefore, 
\begin{equation}
\label{eq:thermo1}
\dot{E}=\frac{dE_1}{dt}+P_{\rm c}\frac{dV_1}{dt},
\end{equation}
governs this region. Here $P_{\rm c}=P_1(R=R_{\rm c})$,  $V_1=4\pi R_{\rm c}^3/3$, and $R_{\rm c}$ is the distance of the contact surface from the center.  The equation of state in this region indicates $E_1=3P_1V_1$.

Region 3 is the pre-shock region with negligible pressure and remains unchanged from the initial conditions given by equation (\ref{eq:rhopro}). 

\section{Self-similar flows}
We apply the similarity transformation to the fluid equations governing flows of the shocked matter in region 2 in the same manner as for the conventional Sedov-Taylor solution.
The profiles of physical quantities are assumed to be described as functions of a non-dimensional independent variable $\xi$,  which can be uniquely determined as
\begin{equation}
\xi=\left(\frac{A}{B}\right)^{1/(\omega-5)}rt^{(3-k)/(\omega-5)}.
\label{eq:xi0}
\end{equation}
Then the shock front radius $R$ and velocity $D$ are described as
\begin{eqnarray}
R(t)&=&\xi_{0}\left(\frac{A}{B}\right)^{1/(5-\omega)}t^{(3-k)/(5-\omega)}, \\
\label{eq:rt}
D(t)&~=&\frac{3-k}{5-\omega}\xi_{0}\left(\frac{A}{B}\right)^{1/(5-\omega)}t^{(\omega-k-2)/(5-\omega)},
\end{eqnarray}
where $\xi_0$ is a constant parameter specifying the shock front. We will use the normalized variable $\lambda$ defined as
\begin{equation}
\lambda=\frac{\xi}{\xi_{0}}=\frac{r}{R}=\frac{1}{\xi_{0}}\left(\frac{A}{B}\right)^{1/(\omega-5)}rt^{(3-k)/(\omega-5)}.
\end{equation}
The shock front is specified by $\lambda=1$.

We transform partial differential equations (\ref{eq:mo})-(\ref{eq:ene}) to ordinary differential equations with respect to $\lambda$. The density, the velocity, and the pressure are assumed to depend on $\lambda$ through non-dimensional functions $\Omega(\lambda)$, $V(\lambda)$, and $\Pi(\lambda)$, respectively:
\begin{eqnarray}
\label{eq:ndimp1}
\rho(r,t)&=&\rho_0\left(R(t)\right)\Omega(\lambda)=B\left[\frac{1}{\xi_{0}}\left(\frac{A}{B}\right)^{1/(\omega-5)}\right]^{\omega}t^{\frac{\omega(3-k)}{\omega-5}}\lambda^{-\omega}\Omega(\lambda), \\
\label{eq:ndimp2}
v(r,t)&=&\frac{r}{t}V(\lambda)=\xi_{0}\left(\frac{A}{B}\right)^{-1/(\omega-5)}\lambda t^{-\frac{3-k}{\omega-k}-1}V(\lambda), \\
\label{eq:ndimp3}
p(r,t)&=&\rho_0\left(R(t)\right)\frac{r^2}{t^2}\Pi(\lambda)=B\left[\frac{1}{\xi_{0}}\left(\frac{A}{B}\right)^{1/(\omega-5)} \right]^{\omega-2}t^{\frac{(\omega-2)(3-k)}{\omega-5}-2}\lambda^{-\omega+2}\Pi(\lambda).
\end{eqnarray}
Using these variables, the fluid equations (\ref{eq:mo})-(\ref{eq:ene}) are rewritten as the following non-dimensional ordinary differential equations.
\begin{equation}
\label{eq:nmoeq}
\left(V-\frac{3-k}{5-\omega}\right)\lambda\dot{V}+V(V-1)-(\omega-2)\frac{\Pi}{\Omega}+\lambda\frac{\dot{\Pi}}{\Omega}=0, 
\end{equation}
\begin{equation}
\label{eq:nconeq}
(3-\omega)V+\lambda\dot{V}+\left(V-\frac{3-k}{5-\omega}\right)\lambda\frac{\dot{\Omega}}{\Omega}=0,
\end{equation}
\begin{eqnarray}
\label{eq:noeneq}
\left(V-\frac{3-k}{5-\omega}\right)\lambda\frac{\dot{\Pi}}{\Pi}-\left(V-\frac{3-k}{5-\omega}\right)\gamma \lambda \frac{\dot{\Omega}}{\Omega}+V(\gamma-1)\omega+2(V-1)=0,
\end{eqnarray}
where the dot means a derivative with respect to $\lambda$. Solving these equations with respect to the derivatives of the non-dimensional functions, we obtain 
\begin{equation}
\label{eq:nomoeq2}
\dot{V}=\frac{\displaystyle \frac{1}{\Omega}[(3\gamma-\omega+2)V-2]-\frac{1}{\Pi}\left(V-\frac{3-k}{5-\omega} \right)\left[V(V-1)-(\omega-2)\frac{\Pi}{\Omega}\right]}{\displaystyle \left(V-\frac{3-k}{5-\omega} \right)^2\frac{\lambda}{\Pi}-\gamma\frac{\lambda}{\Omega}},
\end{equation}
\begin{equation}
\label{eq:nconeq2}
\dot{\Omega}=\frac{V(V-1)\frac{\Omega}{\Pi}-(\omega-2)}{\left(V-\frac{3-k}{5-\omega} \right)^2\frac{\lambda}{\Pi}-\gamma\frac{\lambda}{\Omega}} 
-\frac{(3\gamma-\omega+2)V-2+(3-\omega)V\left[\left(V-\frac{3-k}{5-\omega}\right)^{2}\frac{\Omega}{\Pi}-\gamma\right]}{\left(V-\frac{3-k}{5-\omega}\right)\left[\left(V-\frac{3-k}{5-\omega} \right)^2\frac{\lambda}{\Pi}-\gamma\frac{\lambda}{\Omega}\right]}
\end{equation}
\begin{equation}
\label{eq:noeneq2}
\dot{\Pi}=\frac{\displaystyle \gamma \left[V(V-1)-(\omega-2)\frac{\Pi}{\Omega} \right]-\left(V-\frac{3-k}{5-\omega} \right)\left[\left(3\gamma-\omega-2 \right)V-2 \right]}{\displaystyle \left(V-\frac{3-k}{5-\omega} \right)^2\frac{\lambda}{\Pi}-\gamma\frac{\lambda}{\Omega}}.
\end{equation}
We use the Runge-Kutta method to numerically integrate these equations.

Boundary conditions to integrate equations (\ref{eq:nomoeq2})-(\ref{eq:noeneq2}) are given at  the shock front ($\lambda=1$) and the contact surface.
The physical quantities satisfy the Rankine-Hugoniot relations for a strong shock given at the front as
\begin{equation}
V(1)~=~\frac{k-3}{\omega-5}\frac{2}{\gamma+1}, ~
\Omega(1)~=~\frac{\gamma+1}{\gamma-1}, ~
\Pi(1)~=~\left(\frac{k-3}{\omega-5}\right)^2\frac{2}{\gamma+1}.
\end{equation}

As indicated by equation (\ref{eq:nconeq2}), the density vanishes or diverges to infinity at the contact surface specified by $V_{\rm c}=(3-k)/(5-\omega)$.  On the other hand, the pressure should take a finite value $P_{\rm c}$ determined by equation (\ref{eq:thermo1}).
This boundary condition in turn yields an expression for the temporal evolution of the pressure at the contact surface (and in region 1) and the parameter to specify the shock position $\xi_0$ introduced in equation (\ref{eq:xi0}) as follows. Since the internal energy of the photon gas  $E_1$ is given by $E_1=3P_{\rm c}V_1$, the pressure  $P_{\rm c}$ at the contact surface is obtained by integrating equation  (\ref{eq:thermo1}) as
\begin{equation}
\label{eq:pc}
P_{\rm c}=C_1R_{c}^{-4}+\frac{(5-\omega)(\gamma_1-1)}{3-k-(5-\omega)(k-1)}\frac{3t^{1-k}A}{4\pi R_{c}^{3}},
\end{equation}
where $C_1$ is a constant.
On the other hand, equation (\ref{eq:ndimp3}) indicates
\begin{equation}
\label{eq:pc2}
P_{\rm c}=\frac{\Pi_{\rm c}}{(\lambda_{\rm c}\xi_0)^{\omega-5}}\frac{At^{1-k}}{R_{c}^{3}}.
\end{equation}
These two pressures must be identical. Thus $C_1=0$ and  the constant $\xi_0$ must satisfy the relation
\begin{equation}\label{eq:xi0}
\xi_0=\frac{1}{\lambda_{\rm c}}\left[\frac{4\pi\left(3-k-(5-\omega)(k-1)\right)\Pi_{\rm c}}{5-\omega}\right]^{\frac{1}{\omega-5}}.
\end{equation}
Since the values of $\lambda_{\rm c}$ and $\Pi_{\rm c}$ in the right hand side of this formula are obtained by integrating equations (\ref{eq:nomoeq2})-(\ref{eq:noeneq2}) from the shock front to the contact surface, this formula determines the value of $\xi_0$.

\section{Allowed Ranges of parameters for self-similar flows}\label{sec:range}
Before presenting results, we consider ranges of parameters $k$ and $\omega$ for which a solution exists. The finite amount of the total energy implies $0\leq k<1$ as already mentioned in \S2. Thus the positive shock front velocity $D>0$ is guaranteed as long as $\omega<5$  from equation (\ref{eq:rt}). 

Another condition to be satisfied by the flow in region 1 is that the pressure gradient at the contact surface should not become infinite. Otherwise the pressure vanishes. The pressure gradient at the contact surface can be written with the corresponding non-dimensional pressure $\Pi$ as
\begin{equation}
\label{eq:gradpc}
\frac{\partial p}{\partial r} \biggl \vert_{r=r_{\rm c}}~\propto~\frac{d}{d \lambda}\left(\lambda^{-\omega+2} \Pi \right)\biggl \vert_{\lambda=\lambda_{\rm c}}=\left(-\omega+2\right)\lambda_{c}^{-\omega+1}\Pi_{c}+\lambda_{c}^{-\omega+2}\dot{\Pi}_{c}, 
\end{equation}
 using equation (\ref{eq:ndimp3}). The subscript c indicates variables at the contact surface.
Here $\dot{\Pi}_{\rm c}$ can be expressed as
\begin{equation}
\label{eq:dotpi_c}
\dot{\Pi}_{c}=-\frac{\Omega_{\rm c}}{\lambda_{\rm c}}\left(\frac{k-3}{\omega-5}\right)\left(\frac{k-3}{\omega-5}-1\right)+(\omega-2)\frac{\Pi_{\rm c}}{\lambda_{\rm c}},
\end{equation}
by substituting the expression $V_{\rm c}=(k-3)/(\omega-5)$ into equation (\ref{eq:noeneq2}). Finally, we obtain
\begin{equation}
\frac{\partial p}{\partial r} \biggl \vert_{r=r_{\rm c}}~\propto~\lambda_{\rm c}^{-\omega+1}\Omega_{\rm c}\left(\frac{k-3}{\omega-5}\right)\left(\frac{\omega-k-2}{\omega-5}\right).
\end{equation}
Since $\Omega$ diverges or vanishes at the contact surface, this equation indicates that the pressure gradient needs to vanish at the contact surface to satisfy the boundary condition. Thus $\Omega_{\rm c}$ should vanish at the contact surface or $\omega-k=2$ needs to hold.  Though the pressure gradient might have a finite value if $\omega-k=2$, we find that numerical solutions with parameters satisfying this relation cannot be described by the self-similar solutions especially in region 2.  Therefore  $\Omega_{\rm c}$ needs to be 0 to satisfy the boundary condition.
Then we derive the derivative $\dot{\Omega}_{\rm c}$ of the non-dimensional  density near the contact surface from equation (\ref{eq:nconeq2}).  Using the Taylor expansion for $V$ around the contact surface where $V+(3-k)/(\omega-5)=0$. We obtain
\begin{equation}
\label{eq:dotome}
\frac{\dot{\Omega}_{\rm c}}{\Omega_{\rm c}}=\frac{\left[\left\{3\gamma-1+\left(1-\gamma\right)k\right\}\omega-2k-4\right]}{\left(\lambda-\lambda_{\rm c}\right)\left[\left(1-k\right)\omega-\left(3\gamma+2\right)\left(3-k\right)+10\right]}.
\end{equation}
Thus the boundary condition at  the contact surface is satisfied if 
\begin{equation}\label{ineq:exist}
\left[\left\{2+\left(\gamma-1\right)\omega\right\}k+\left(1-3\gamma\right)\omega+4\right]\left\{\left(3\gamma-\omega+2\right)k-9\gamma+\omega+4\right\} \leq 0.
\end{equation}
This can be rewritten as the bound of the value of $k$ as
\begin{equation}\label{ineq:existk}
\frac{\left(3\gamma-1\right)\omega-4}{2+\left(\gamma-1\right)\omega}\leq k\leq\frac{9\gamma-\omega-4}{3\gamma-\omega+2}.
\end{equation}
Here we have limited the range of the adiabatic index $\gamma$ from 1 to 2 to simplify the argument. The value of $k$ has been already limited to $0\leq k<1$ for the flow to have a finite total energy and the value of $\omega$ is limited to $\omega<5$ for the shock to move outward. There exists $k$ meeting these three criteria if $\omega<3/\gamma$.

\section{Results}
In this section, we present self-similar solutions for several parameter sets. At the same time, we show the corresponding solutions obtained by directly integrating fluid equations (\ref{eq:mo})-(\ref{eq:ene}) for comparison to check whether the self-similar solutions actually describe real flows. We use the Harten-Lax-van Leer Contact (HLLC) scheme \citep{Toro1994} for solving numerical flux and the Monotonic Upstream-Centered Scheme for Conservation Laws (MUSCL) scheme \citep{Leer1977} to improve the spatial resolution. 
The calculated region is divided into $N$ equidistant spherical shells.The boundary conditions are imposed as zero gradient conditions except for the velocity at the center, which is set to zero.  Additionally, we investigate the flows resulting from  parameters for which our solutions cannot satisfy the boundary conditions. 

\begin{figure}[t]
\begin{center}
\includegraphics[clip,width=10.0cm]{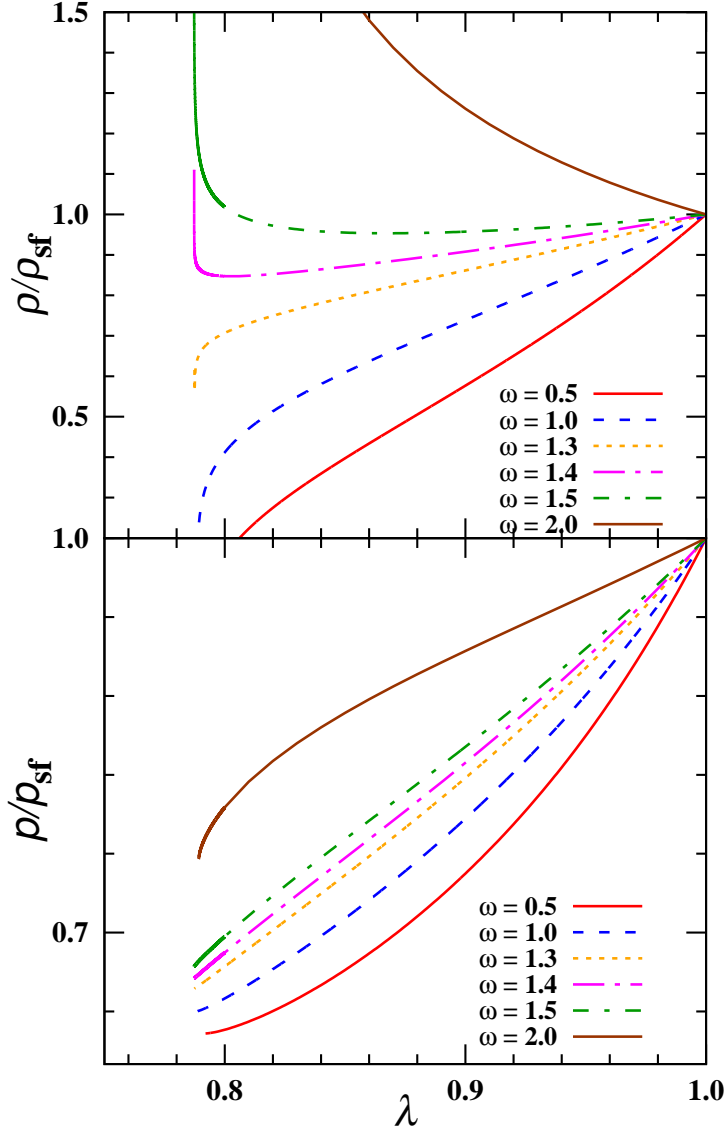}
\end{center}
 \caption{Density (top panel) and pressure (bottom panel) profiles of self-similar solutions with $k=0.5$ and several values of $\omega$ indicated in the panels. The profiles are shown as functions of the non-dimensional variable $\lambda=r/R$. }\label{fig:2}
\end{figure}

Figure \ref{fig:2} shows the pressure and density profiles for self-similar solutions with $k=0.5$ and several values of $\omega$.  The adiabatic index is fixed to $\gamma=5/3$. Though the pressure at the contact surface apparently has a finite value due to the coarse grid in this figure, the analytical expression for the self-similar flow near the contact surface suggests that  the pressure diverges at the contact surface if $\omega \ge 1.4$ for $k=0.5$ according to the inequality (\ref{ineq:exist}).

\begin{figure}[t]
\begin{center}
\includegraphics[clip,width=18.0cm]{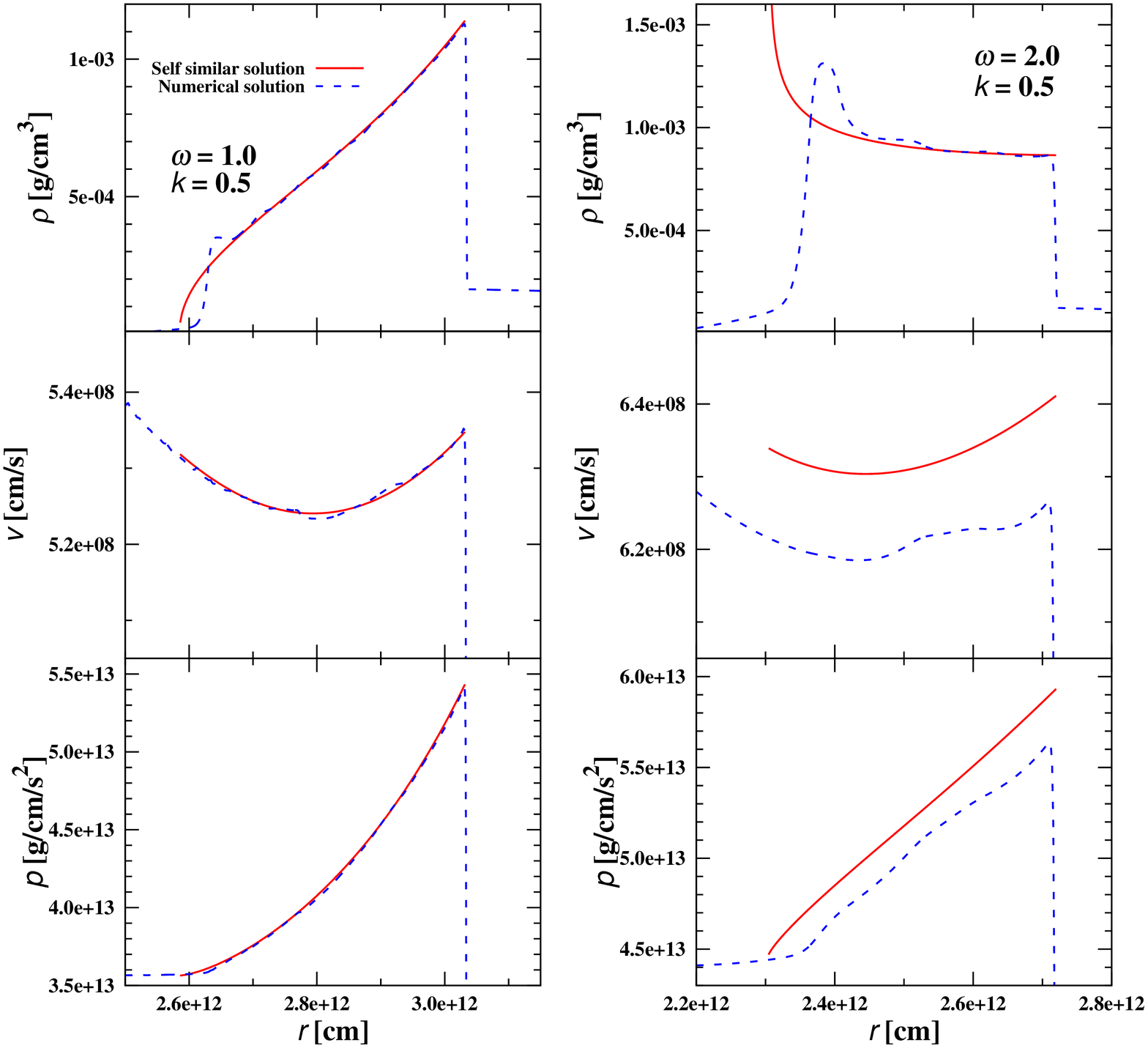}

  \end{center}
    \caption{Density (top), velocity (middle), and pressure (bottom) profiles of self-similar solutions (solid lines) and numerical solutions (dashed lines) as functions of the radius.   The left panels show the solutions for the parameters $\omega=1.0$, $k=0.5$, $A=1.58\times10^{50}$ (in cgs units), and $B=2.21\times10^{-4}$ (in cgs units).
  The right panels show the solutions with parameters $\omega=2.0$ and $k=0.5$ for which the self-similar shocked flow does not satisfy the boundary conditions. See the text for more details.
  }  \label{fig:3}
\end{figure}

Figures \ref{fig:3} compare self-similar solutions and the corresponding numerical solutions. The adiabatic index of the flow is assumed to be $\gamma=4/3$.
 Here we integrate fluid equations (\ref{eq:mo})-(\ref{eq:ene}) with a central energy source described by equation (\ref{eq:eneinject}). The initial density distribution is given by equation (\ref{eq:rhopro}). The initial pressure distribution is assumed to be uniform and  set to 0.1  erg cm$^{-3}$. The flows are calculated in a region with $R=3.5\times10^{12}~\rm cm$, the cell size is $\Delta r=R/N=1.75\times10^9~\rm cm$. The left panels show the profiles resulting from  an energy source with $k=0.5$ and the initial density with $\omega=1$ for which a self-similar solution exists. The values of the other parameters are set to $A=E(1-k)\tau^{k-1}=1.58\times10^{50}$ and $B=M(3-\omega)R^{\omega-3}/4\pi=2.21\times10^{-4}$ in cgs units, where the total mass $M=20$ $\rm{M_\odot}$, the amount of injection energy $E=10^{51}~\rm erg$ for $\tau=10$ s. The right panels show the profiles with $k=0.5$ and $\omega=2.0$, $B=8.63\times10^{20}$ and the same value of $A$ used for $\omega=1$. Though the non-dimensional quantities can be integrated for this set of parameters ($k=0.5$,~$\omega=2.0$), we cannot specify the value of $\xi_0$ from equation (\ref{eq:xi0}) because this flow does not satisfy the boundary condition at the contact surface. To compare it with a result from numerical integration of fluid equations, we decide the value of $\xi_0$ to align the positions of the shock fronts in these two solutions at $t=3.03\times10^{3}$ s.  The deviations of the velocity and the pressure from the numerical solutions in the right panels clearly indicate that the self-similar solution cannot yield a self-consistent flow.

\section{Conclusions and discussion} \label{sec:sum}

We calculate blast waves driven by a central energy source with the time-dependent energy injection rate given by equation (\ref{eq:eneinject}). The wave propagates in the ambient matter with a density profile described as equation (\ref{eq:rhopro}). The system is divided into three regions.
We limit the range of parameters $\omega $ and $k$ for which self-similar solutions exist. At first,  a finite explosion energy restricts the value of $k$ to $0\leq k<1$. Second, since the wave propagates in the positive radial direction, $\omega<5$. The boundary condition at the contact surface gives further constraints as discussed in section \ref{sec:range}. 

Our self-similar solutions are confirmed to describe flows resulting from a central energy source by comparison with numerical solutions.  
We also  investigate what occurs when we chose the parameter set for which the self-similar solution does not exist. To specify the parameter $\xi_0$,  which is supposed to be determined from the boundary condition at the contact surface, we align the locations of the shock fronts at a certain time. Then the self-similar solution yields velocities and pressures significantly higher than the numerical solution. This indicates that the self-similar solution cannot be self-consistent for this set of parameters.

As \citet{Ostriker1988} pointed out, the present solutions with continual energy injection are applicable to the blast wave formed by a stellar wind from a massive star in uniform interstellar matter. The parameters for this situation $\omega=0$ and $k=0$ are within the allowed range we have investigated in Section 4.
On the other hand, we cannot use our solutions to describe the blast wave propagating in the stellar mantle as a result of energy injection from a central engine realized in SN explosions. For example, the energy injection from magnetic dipole radiation by a magnetar ($k=0$) into the  envelope of a massive star ($\omega>9/4$) yields parameters outside the applicable range.

\bibliographystyle{apj}
\bibliography{reference}

\end{document}